\documentclass[pra,amssymb]{revtex4}

\usepackage{appendix}
\usepackage{graphicx}
\usepackage{amssymb,amsmath} 

\usepackage[polish,english]{babel}
\usepackage[T1]{fontenc}

\def\B#1{\left(#1\right)}
\def\BB#1{\left[#1\right]}

\def\be{\begin{equation}}
\def\ee{\end{equation}}

\def\bee{\begin{equation*}}
\def\eee{\end{equation*}}

\def\for{\ \ {\rm for} \  }

\def\kneq{k\setminus\{0\}} 
\def\kkneq{k\setminus\{0,\pi \}} 
\def\kkkneq{k\setminus\{\pi\}} 

\begin{document}

\title{Exact results for fidelity susceptibility of the quantum Ising model: 
The interplay between  parity, system size, and magnetic field}

\author{Bogdan Damski$^1$ and Marek M. Rams$^{2,3}$} 
\affiliation{$^1$\mbox{ Institute of Physics, Jagiellonian University, Reymonta 4, 30-059 Krak\'ow, Poland} \\
$^2$\mbox{Vienna Center for Quantum Science and Technology, Faculty of Physics, University of Vienna, Vienna, Austria}\\
$^3$\mbox{Institute of Physics, Krak\'ow University of Technology, Podchor\k{a}\.zych 2, 30-084 Krak\'ow, Poland}
}
\begin{abstract}
We derive an exact closed-form expression  for fidelity susceptibility of even- and odd-sized 
 quantum Ising chains in a transverse field. 
To this aim, we diagonalize the Ising Hamiltonian and study the gap between its  positive and 
negative parity subspaces. We  derive an exact closed-form expression for the
gap and use it to identify the parity of the  ground state. 
We point out  misunderstanding in some of the former studies of
fidelity susceptibility  and discuss its  consequences. Last but not
least,  we  rigorously analyze the properties of the gap. For example,  we
derive analytical expressions showing its exponential dependence on the ratio
between the system size and the correlation length. 
\end{abstract}
\maketitle

\section{Introduction}
Quantum phase transitions happen  at zero absolute temperature when there are
competing interactions trying to order the sample in different ways \cite{Sachdev}. 
The balance between them is typically controlled by an external field 
such as  a magnetic field 
acting on spins or the laser field imposing a periodic
potential on  cold  atoms or ions \cite{SachdevToday,LewensteinAdv}. 
When this field reaches the critical value, the system is at the quantum critical point.
Then, a small tilt of the field moves the system into one of the phases changing its
properties significantly (e.g. magnetization, correlation functions, etc.).

Several approaches to quantum phase transition
have been proposed. The fairly recent one that we explore 
focuses on the (ground state) fidelity: 
$\left|\langle\psi(g)|\psi(g+\delta)\rangle\right|$,
where $|\psi(g)\rangle$ is a ground-state wave-function of some 
Hamiltonian $\hat H(g)$, $g$ is the external field whose variation 
induces a quantum phase transition, and  $\delta$ is a small, but otherwise
arbitrary,  shift of this field \cite{Zanardi}.  
This approach has been studied in a stunning variety of models. 
For example, numerous spin models (Ising, XY, XXZ, Heisenberg, Kitaev, Lipkin-Meshkov-Glick, etc.),  
Hubbard, Bose-Hubbard, and Hubbard-Holstein models, Harper model, Dicke model, Luttinger liquid model, etc. 
(see the review paper of Gu \cite{GuReview} and references therein).
Dramatic change of the system's properties 
across the critical point results in a drop of fidelity
enabling both the location of the critical point and the determination of the 
universal critical  exponent $\nu$ characterizing the divergence of the 
correlation length  \cite{Zanardi,Chen2008,ABQ2010,Polkovnikov,MarekBodzioPRL,SenPRB2012,Zhou,Zhou1,Adamski2013,Polk1}.

We will study here fidelity susceptibility $\chi(g)$, which is defined 
through the Taylor expansion of fidelity in the field shift $\delta$:
\be
\left|\langle\psi(g)|\psi(g+\delta)\rangle\right|=
1-\chi(g)\frac{\delta^2}{2}-\chi'(g)\frac{\delta^3}{4}+ {\cal O}\B{\delta^4},
\label{mnbvcxz}
\ee
where we assumed that the ground states are normalized: $\langle\psi(g)|\psi(g)\rangle=1$. 
This expansion is valid in the limit of $\delta\to0$ taken at the fixed system size $N$. Fidelity 
 approaches  unity in this limit. 
 Note that 
there exists the other limit of $N\to\infty$  taken at the fixed field shift
$\delta$ \cite{MarekBodzioPRL,SenPRB2012,Zhou,Zhou1,Adamski2013}. In such
limit fidelity approaches zero, which 
is known as the Anderson orthogonality  catastrophe \cite{Anderson1967}.

\section{Model}
\label{sec_model}
We study fidelity susceptibility of the quantum Ising model in a
transverse field. 
This is a paradigmatic model of  a quantum phase transition \cite{Sachdev}
that can be experimentally realized in magnetic materials such as  CoNb$_2$O$_6$
\cite{Coldea}. Its promising experimental emulation in an ion chain was proposed 
in Ref. \cite{ion_sim}.
The Hamiltonian that we study reads
\be
\hat H(g) = -\sum_{i=1}^N \left(\sigma^x_i\sigma^x_{i+1} + g\sigma^z_i\right),
\label{HIsing}
\ee
where $g$ is the  magnetic field,  $N>2$ is the  number of spins, and we
assume periodic boundary conditions $\vec\sigma_{N+1}=\vec\sigma_1$.
For $|g|>1$ the system is in the paramagnetic phase, 
while for $|g|<1$ it is in the
ferromagnetic phase. The critical points are at  $g_c=\pm1$.

A simple calculation leads to the following expression for fidelity
susceptibility \cite{Zanardi}
\be
\chi(g)=\frac{1}{4}\sum_{k>0} \frac{\sin^2k}{\B{g^2-2g\cos k+1}^2},
\label{chi}
\ee
where $k$ is the momentum of quasiparticles introduced to
diagonalize the Hamiltonian (\ref{HIsing}). 

Our  goal  is to  derive an exact closed-form expression for fidelity
susceptibility for {\it all} system sizes $N$ and magnetic fields $g$
(for some approximate studies of fidelity susceptibility in the Ising chain see e.g. Refs.
\cite{Zanardi,GuReview,Chen2008}).
The first step in this direction was  done in Ref.  \cite{BDfid3}, 
where even-sized chains were studied. We extend this result to the odd system sizes
here. This requires precise determination of what momenta 
$k$ should be placed into the sum (\ref{chi}), which we discuss 
in Secs. \ref{sec_diagonalization} and \ref{sec_ground}. 
The two choices that were employed in the past 
are  
\be
k=\frac{\pi}{N}, \frac{3\pi}{N}, \frac{5\pi}{N},\dots
\label{kpos}
\ee
and
\be
k=\frac{2\pi}{N}, \frac{4\pi}{N},\frac{6\pi}{N},\dots
\label{kneg}
\ee
We will  {\it analytically} show in Secs. \ref{sec_fidelity} and
\ref{sec_discussion}
that these two sets of momenta lead to significantly different results for fidelity
susceptibility near the critical point (this difference was numerically pointed out 
in the context of Quantum Monte Carlo simulations of fidelity susceptibility in Ref. \cite{ABQ2010}). 
An exact closed-form expression for fidelity susceptibility  in odd-sized systems will be derived in Sec.
\ref{sec_fidelity} and combined with the even $N$ result from Ref.
\cite{BDfid3}, so that the complete expression for fidelity susceptibility in the Ising
chain will be finally obtained. Finally, we will discuss in Sec.
\ref{sec_discussion}  the consequences of incorrect quantization of momenta.

\section{Diagonalization}
\label{sec_diagonalization}
All the steps discussed in this section are very elementary, but they are  often presented incompletely in 
the literature. This presumably leads to misunderstandings discussed in Sec.
\ref{sec_discussion}. As a matter of fact, the discussion of the ground state of the Hamiltonian 
(\ref{HIsing}) for all magnetic fields and system sizes is not as trivial as usually assumed
(Sec. \ref{sec_ground}).
We mention in passing that none of the standard references to the Ising model
diagonalization \cite{Lieb1961,Katsura1962,Pfeuty,Sachdev} provides
a complete discussion of the ground state determination in a {\it finite} system. 

The first step to diagonalize the Hamiltonian
(\ref{HIsing})  is to perform the Jordan-Wigner transformation:
\bee
\sigma_i^z=1-2\hat c_i^\dag \hat c_i, 
\ \sigma_i^x=\B{\hat c_i + \hat c_i^\dag}\prod_{j<i}\B{1-2\hat c_j^\dag\hat c_j}, 
\  \{\hat c_i,\hat c_j^\dag\} = \delta_{ij}, 
\ \{\hat c_i,\hat c_j\} = 0.
\eee
After this transformation and some additional manipulations one finds
\bee
\hat H = -\sum_{i=1}^{N-1} \hat f_{i,i+1} +\hat f_{N,1} \hat P - g \sum_{i=1}^N (\hat c_i \hat c_i^\dag  - \hat c_i^\dag \hat c_i),
\ \hat f_{i,j} = \hat c_i^\dag\hat c_j - \hat c_i \hat c_j^\dag - \hat c_i\hat c_j +  \hat c_{i}^\dag \hat c_j^\dag,
\ \hat P =  \prod_{i=1}^N \B{1-2\hat c_i^\dag \hat c_i},
\eee
where $\hat P$ is the parity operator, whose eigenvalues are either $+1$
(positive parity eigenstate) or $-1$ (negative parity eigenstate).
This  can be rewritten as 
\bee
\hat H = \hat H^+ \hat P^+ + \hat H^{-} \hat P^{-}, 
\ \hat H^\pm = -\sum_{i=1}^{N} \BB{\hat f_{i,i+1} + g\B{\hat c_i \hat c_i^\dag - \hat c_i^\dag \hat c_i}},
\ \hat P^\pm = \frac{1}{2} \B{1\pm\hat P},
\eee
where $\hat P^+$ and $\hat P^-$ are the projectors onto the subspaces with an even and odd
number of $c$-particles, respectively. The boundary conditions for $\hat H^+$ 
are antiperiodic
\be
\hat c_{N+1}=-\hat c_1,
\label{canty}
\ee
while for $\hat H^-$ they are periodic
\be
\hat c_{N+1}=\hat c_1.
\label{cper}
\ee

Since the parity operator commutes with the Hamiltonian,
one can find the eigenstates of both $\hat H$ and $\hat P$.
This allows for  independent diagonalization in the subspaces of
positive and negative  parity. 
These simple calculations have to be done
independently in even- and odd-sized chains.

{\bf Even number of spins, positive parity subspace.}
We perform the Fourier transform
\be
\hat c_j =\frac{\exp\B{-i\pi/4}}{\sqrt{N}}\sum_k\hat c_k \exp\B{ikj},
\label{fourier}
\ee
where 
\bee
k = \pm\frac{\pi}{N},\pm\frac{3\pi}{N},\dots,\pm\B{\pi-\frac{\pi}{N}}
\eee
to satisfy Eq. (\ref{canty}).
This leads to  the Hamiltonian
\bee
\hat H^+({\rm even}\  N) = \sum_k \hat h_k, \ \hat h_k = [g-\cos(k)][\hat c_k^\dag \hat c_k -\hat c_{-k} \hat c_{-k}^\dag  ]+ \sin(k) [\hat c_k^\dag\hat c_{-k}^\dag + \hat c_{-k}\hat c_{k}],
\eee
which is diagonalized by the Bogolubov transformation 
\be
\hat c_k = \cos\B{\frac{\theta_k}{2}}\hat\gamma_k - \sin\B{\frac{\theta_k}{2}}\hat\gamma_{-k}^\dag, 
\ \B{\sin \theta_k,\cos \theta_k} = \B{\frac{\sin(k)}{\epsilon_k}, \frac{g-\cos(k)}{{\epsilon_k}}},
\label{bogolubov}
\ee
where another quasiparticle operator, 
\bee
\hat\gamma_k =\cos\B{\frac{\theta_k}{2}}\hat c_k +\sin\B{\frac{\theta_k}{2}}\hat c_{-k}^\dag, \
\{\hat\gamma_k,\hat\gamma_{k'}^\dag\}=\delta_{kk'}, \
\{\hat\gamma_k,\hat\gamma_{k'}\}=0,
\eee
has been introduced. Finally, the diagonalized Hamiltonian becomes 
\bee
\hat H^+({\rm even} \  N)=\sum_k \epsilon_k \B{2 \hat\gamma_k^\dag\hat\gamma_k-1}, \
\epsilon_k = \sqrt{[g-\cos(k)]^2+\sin^2(k)}.
\eee
The ground-state is then  annihilated by all the operators
$\hat\gamma_k$. It can be written as 
\be
\prod_{k>0} \B{\cos\B{\frac{\theta_k}{2}} -
\sin\B{\frac{\theta_k}{2}}\hat c_k^\dag\hat c_{-k}^\dag}|{\rm vac}\rangle,
\label{wavefunction}
\ee
where $|{\rm vac}\rangle$ is annihilated by all $\hat c_k$ operators.
This ground state  contains an even number of 
$c$-particles. It belongs to the positive parity subspace.
Its eigenenergy is
\be
\varepsilon^{+}({\rm even} \ N)= -\sum_{k}\epsilon_k.
\label{en1}
\ee
Combining Eqs. (\ref{wavefunction}), (\ref{bogolubov}), and 
(\ref{mnbvcxz}) one  obtains Eq. (\ref{chi}). The same summand  appears for all parities and system sizes.

{\bf Even number of spins, negative parity subspace.} 
We perform the Fourier transform (\ref{fourier}) with momenta 
\bee
k=0,\pm\frac{2\pi}{N},\pm\frac{4\pi}{N},\dots,\pm\B{\pi-\frac{2\pi}{N}},\pi
\eee
to ensure that Eq. (\ref{cper}) holds.
The transformed Hamiltonian reads:
\bee
\hat H^-({\rm even}\  N) = \sum_{\kkneq} \hat h_k + (g-1)(\hat c_0^\dag \hat c_0-\hat c_0 \hat c_0 ^\dag) 
+(g+1)(\hat c_{\pi}^\dag \hat c_{\pi}- \hat c_{\pi}\hat c_{\pi}^\dag ).
\eee
The $\kkneq$ part of the Hamiltonian is diagonalized with the Bogolubov
transformation (\ref{bogolubov}). Thus, there is an even number of quasiparticles
in those modes [see the wave-function (\ref{wavefunction})]. 
Therefore, the negative parity ground state will have either the
$k=0$ or the $k=\pi$ mode excited. Minimizing the
eigenenergy we find that the $k=0$ mode is occupied, the $k=\pi$ one is empty, and the eigenenergy is 
\be
\varepsilon^-({\rm even} \ N) = -\sum_{\kkneq} \epsilon_k-2.
\label{en2}
\ee

{\bf Odd number of spins, positive parity subspace.} 
We perform the Fourier transform (\ref{fourier}) with momenta 
\bee
k=\pm\frac{\pi}{N},\pm\frac{3\pi}{N},\dots,\pm\B{\pi-\frac{2\pi}{N}},\pi
\eee
to ensure that Eq. (\ref{canty}) holds.
The transformed Hamiltonian reads:
\bee
\hat H^+({\rm odd}\  N) = \sum_{\kkkneq} \hat h_k  + (g+1)(\hat c_{\pi}^\dag \hat c_{\pi}-\hat c_{\pi} \hat c_{\pi}^\dag).
\eee
The $\kkkneq$ part of the Hamiltonian is diagonalized with the Bogolubov
transformation (\ref{bogolubov}). Thus, there is an even number of quasiparticles
in those modes. Therefore, the positive  parity ground state will have to have the
$k=\pi$ mode empty. Its eigenenergy is
\be
\varepsilon^+({\rm odd} \ N) = -\sum_{\kkkneq}\epsilon_k-g-1.
\label{en3}
\ee

{\bf Odd number of spins, negative parity subspace.} 
We perform the Fourier transform (\ref{fourier}), but this time the summation goes over 
momenta 
\bee
k=0,\pm\frac{2\pi}{N},\pm\frac{4\pi}{N},\dots,\pm\B{\pi-\frac{\pi}{N}}
\eee
to satisfy Eq. (\ref{cper}). 
After this transform
\bee
\begin{aligned}
&\hat H^-({\rm odd}\  N) = \sum_{\kneq} \hat h_k  + (g-1)(\hat c_0^\dag \hat c_0 - \hat c_0 \hat c_0^\dag).
\end{aligned}
\eee
The $\kneq$ part of the Hamiltonian is diagonalized 
by the Bogolubov transformation (\ref{bogolubov}).
This implies that the number of quasiparticles in the $\kneq$ modes is even. The
total number of quasiparticles in the negative parity ground state is odd.
Therefore, the diagonal $k=0$ mode has to be occupied. The eigenenergy of such
a ground state is 
\be
\varepsilon^-({\rm odd} \ N) = -\sum_{\kneq} \epsilon_k+g-1.
\label{en4}
\ee

Next, we will compare the ground state energies in the positive and negative parity  
subspaces to find out the ground state of the whole (not just the {\it parity-projected})
Hamiltonian.

\section{Ground state and gap}
\label{sec_ground}
We have identified ground states in the positive and negative parity subspaces.
Next, we find that   
\begin{align}
&\varepsilon^- - \varepsilon^+  =  
g^N \int_0^1 dt\, \frac{4 N}{\pi} \frac{t^{N-3/2} \sqrt{(1-t)(1-g^2 t)} }{1-(gt)^{2N} } \ \ {\rm for} \ \ |g|<1,
\label{gap} \\
&\varepsilon^- - \varepsilon^+ ={\rm sign}\B{g^N} \B{2|g|-2} +   g^{-N} \int_0^1 dt\,\frac{4 N}{\pi} \frac{t^{N-3/2} \sqrt{(1-t)(g^2-t)} }{1- t^{2N}/g^{2N}} \ \ {\rm for} \ \ |g|>1,
\label{gappara}\\
&\varepsilon^- - \varepsilon^+ = 2\tan\B{\frac{\pi}{4N}} {\rm sign}\B{g^N}\ \ {\rm at} \ \ g=\pm1.
\label{gapcritical}
\end{align}
Expressions (\ref{gap}) -- (\ref{gapcritical}) are valid for any system size $N>2$. 
Two remarks are presented below.

First, these expressions show that 
\bee
{\rm sign} \B{\varepsilon^- - \varepsilon^+} = {\rm sign}\B{g^N},
\eee
which holds for all magnetic fields $g$ and system sizes $N$. Therefore,
the ground state of the Hamiltonian (\ref{HIsing}) has positive parity
when $N$ is even. On the other hand, when $N$ is odd, the parity of the 
ground state is positive for $g>0$ and negative for $g<0$.
This has interesting consequences for fidelity susceptibility (Sec. \ref{sec_fidelity}). These
remarks are summarized in Table \ref{table}; see also Ref. \cite{ABQ2010},
where the Perron-Frobenius theorem is employed to argue that 
the parity of the ground state is positive for $g>0$.

Second, the above expressions quantify the gap between positive and negative
parity ground states, which is important in  understanding symmetry breaking
in the Ising model, 
as well as in other contexts including 
adiabatic driving across the first order phase transition \cite{Sondhi2012}. 
The gap in the ferromagnetic phase is  bounded by the following inequalities
\be
\max \B{ g^N \frac{2}{\sqrt{\pi}}\frac{\sqrt{1-g^2}}{\sqrt{N}} , g^{N} \frac{4g}{\pi N}}  \le\varepsilon^- - \varepsilon^+ \le g^N 2 \frac{\sqrt{1 - g^2}}{\sqrt{N-1}} + g^{N} \frac{\pi g}{2N-1},
\label{smallg}
\ee
where equality happens at $g=0$ (see Appendix \ref{appendix_gap} for details;
we assumed here that $0\le g\le1$ for convenience).
Written in such a way, inequality (\ref{smallg}) shows two limits. 

In the thermodynamic limit -- i.e., when the correlation length in the infinite Ising chain 
\cite{BarouchMcCoy1971}
\bee
\xi \sim 1/|\ln g|
\eee
is much smaller 
than the size of the system -- we see from inequality (\ref{smallg}) that 
the gap is exponentially small in $N$: 
\bee
\varepsilon^--\varepsilon^+ = {\cal O}\B{\exp\B{-N/\xi}/\sqrt{N}}. 
\eee
This result is in agreement with the estimation from Ref. \cite{Sondhi2012},
where it is stated that the gap is smaller than any power law in $N$.
It quantitatively shows how the two fold 
degeneracy of the ground state emerges when the system size increases.

When the size of the system is smaller than the correlation length,  inequality (\ref{smallg}) shows that 
\bee
\varepsilon^- - \varepsilon^+ = {\cal O}\B{1/N}.
\eee
The gap disappears then in the ferromagnetic phase as if the system would be at 
the critical point (\ref{gapcritical}).
It is so because this is the limit where finite-system size effects dominate
physics. 

We mention also that one can use Eq. 
(\ref{gap})  to prove  that the gap is monotonically increasing from $g=0$ to $g=\pm1$,
which is discussed in Appendix \ref{appendix_gap}. Finally, we notice that the gap is always 
``macroscopic'' on the paramagnetic side (\ref{gappara}) -- the symmetry breaking does not occur there.

The qualitative content of these remarks about the disappearance of the gap 
is, of course, discussed in standard textbooks such
as Ref. \cite{Sachdev}. The novelty here is that we managed to obtain the
closed-form expression for the gap, which allowed us to rigorously quantify its behavior. 
This adds another piece of the exact analytical insight  to the plethora of already known
analytical results for the quantum Ising model.

We will now discuss derivation of Eqs. (\ref{gap}) -- (\ref{gapcritical}). We
start from writing  $\epsilon_k$ as $\sum_{\ell=0}^{\infty} a_\ell \cos\B{\ell k}$. 
Next, we substitute this expansion into Eqs. (\ref{en1}) -- (\ref{en4}) and 
use the identity $\sum_{s=0}^{n-1}\cos(x+sy)=\cos(x+(n-1)y/2)\sin(ny/2){\rm cosec}(y/2)$ 
to obtain
\be
\varepsilon^- - \varepsilon^+ = {\rm sign}\B{g^N} \Theta\B{|g|-1}\B{2|g|-2}-2N\B{a_N + a_{3N}+a_{5N}+\dots},
\label{gapfourier}
\ee
where $\Theta$ is the Heaviside step function. This expression holds for any
system size $N>2$ and any magnetic field $g$.

\begin{table}
\begin{tabular}{c|c|c}
             & $g>0$ & $g < 0$ \\ \hline
even $N$ & +   & + \\ \hline
odd $N$ & +   & -- \\ \hline
\end{tabular}
\caption{Parity of the Ising chain ground state.}
\label{table}
\end{table}

Subsequently, we assume that $0<g<1$ 
(the derivation for different values of $g$ proceeds in almost identical way and we 
will only quote the final result later). We write the Fourier coefficients for $\ell>0$ as
\bee
a_\ell = \frac{1}{\pi} \int_{-\pi}^{\pi} dk\,\cos(k\ell) \sqrt{[g-\cos(k)]^2+\sin^2(k)}.
\eee
Introducing $z=e^{ik}$, it is equivalent to
\bee
a_\ell = -\frac{1}{\pi} \oint_{|z|=1}dz\,\sqrt{\frac{(z-g)(g z-1)}{z}}z^{\ell-1},
\eee
with branch cuts along $(0,g) \cup (1/g,\infty)$.
By deforming the integration contour, this leads to
\be
a_\ell = -\frac{2g^\ell}{\pi} \int_0^1dt\,\sqrt{(1-t)(1-g^2t)}\,t^{\ell-3/2}. 
\label{al}
\ee
Substituting Eq. (\ref{al})  into Eq. (\ref{gapfourier}), one obtains Eq.
(\ref{gap}). The derivation for $-1<g<0$ involves slightly different branch cuts and leads to the same 
final result. Thus, Eq.  (\ref{gap}) holds for any $|g|<1$.

The gap in the paramagnetic phase can be obtained through the Kramers-Wannier duality
mapping   
\be
g\leftrightarrow\frac{1}{g}
\label{KW}
\ee
from the gap in the ferromagnetic phase. This duality symmetry follows from 
the symmetry of the classical two dimensional Ising model discussed in the
seminal Ref. \cite{Kramers1941}. The map between the two dimensional classical
model and the one dimensional quantum model that we consider is described in Ref. \cite{Suzuki1976}.
The basics of the duality symmetry in the quantum context are discussed in 
Ref. \cite{Kramers}, while its implications for fidelity and fidelity
susceptibility are discussed in Refs. \cite{Zhou} and \cite{BDfid3}, respectively. 

For the gap, we find that  
\be
\varepsilon^-(g)-\varepsilon^+(g) = {\rm sign}\B{g^N}\B{2|g|-2} +
|g|\BB{\varepsilon^-\B{\frac{1}{g}} - \varepsilon^+\B{\frac{1}{g}}} \for |g|>1.
\label{dual}
\ee
One can show this by performing the mapping  (\ref{KW}) on Eqs.  (\ref{en1}), (\ref{en2}), (\ref{en3}) and 
(\ref{en4}), and  then noting that  $\epsilon_k(1/g)=\epsilon_k(g)/|g|$. Combining
Eqs. (\ref{dual}) and (\ref{gap}) one obtains Eq. (\ref{gappara}).

Eq. (\ref{gapcritical})  can be directly evaluated with   
$\sum_{s=0}^{n-1}\sin(x+sy)=\sin(x+(n-1)y/2)\sin(ny/2){\rm cosec}(y/2)$. 
The same result can be obtained by taking the limit  of $g\to\pm1$ 
in Eqs.  (\ref{gap}) and (\ref{gappara}).

Finally, we mention that it would be interesting to apply 
the above  summation technique to compute 
the gap between the positive and negative 
parity ground states of the XY model. 
This gap was studied in Ref. \cite{XY},
but no exact  closed-form expressions were derived.
It was found numerically in Ref. \cite{XY} that the
parity of the ground state changes as the magnetic field is varied. This happens 
in the XY model for both even- and odd-sized systems at different magnetic
fields. It would be quite interesting to provide an analytical
characterization of this phenomenon coined in Ref. \cite{XY} as vacua
competition.

\section{Fidelity susceptibility}
\label{sec_fidelity}
We can finally focus on our main goal: The  computation of fidelity
susceptibility.
Using the results from Sec. \ref{sec_diagonalization}, we find that for any $N>2$ and any magnetic field $g$
fidelity susceptibility of the lowest energy eigenstate in the positive parity subspace is 
\be
\chi^{+}(g)= 
\frac{N^2}{16g^2}\frac{g^N}{\B{g^N+1}^2}+\frac{N}{16g^2}
\frac{g^N-g^2}{\B{g^N+1}\B{g^2-1}},
\label{chipositive}
\ee
while in the negative parity subspace it is 
\be
\chi^{-}(g)= 
-\frac{N^2}{16g^2}\frac{g^N}{\B{g^N-1}^2}+\frac{N}{16g^2}
\frac{g^N+g^2}{\B{g^N-1}\B{g^2-1}}.
\label{chinegative}
\ee

Thus, employing the results of Sec. \ref{sec_ground}, we find that 
fidelity susceptibility of the {\it ground state} of the  Ising
Hamiltonian is the following. For even system sizes $\chi(g)=\chi^{+}(g)$
for any $g$, while for odd system sizes $\chi(g)=\chi^+(g)$ for $g>0$ and 
$\chi(g)=\chi^-(g)$ for $g<0$ (Table \ref{table}). This can be written as
\bee
\chi(g)= 
\frac{N^2}{16g^2}\frac{|g|^N}{\B{|g|^N+1}^2}+\frac{N}{16g^2}
\frac{|g|^N-g^2}{\B{|g|^N+1}\B{g^2-1}}
\eee
for any system size $N>2$ and any magnetic field $g$. This result provides the
ultimate expression for fidelity susceptibility of the Ising chain in a
transverse field. It is
 worth to stress that its derivation required not only the ability to 
compute the sum (\ref{chi}), but also the careful diagonalization
of the Ising Hamiltonian followed by the rigorous analysis of the gap between 
its positive and negative parity subspaces. We also mention that the
difference between fidelity susceptibility of the positive and negative parity
ground states was numerically quantified in Ref. \cite{ABQ2010}.

We will now discuss the key steps allowing for the  derivation of Eqs.
(\ref{chipositive}) and (\ref{chinegative}) for $g>0$. The results for $g<0$
can be obtained from the $g>0$ ones by changing the summation in Eq. (\ref{chi}) from $k$ to $\pi-k$.

The positive parity result
(\ref{chipositive}) for even $N$ was obtained in Ref. \cite{BDfid3}. Since an analogical 
calculation can be performed for odd $N$ and its outcome is exactly the same, we will not 
repeat it here.

Thus, we focus on the derivation of  the negative parity result, i.e., Eq.
(\ref{chinegative}). Since the calculations for even and odd $N$ are similar
and lead to the same final result,  Eq. (\ref{chinegative}), we will sketch
the
derivation of the odd $N$ result only. Our calculation starts from the identity \cite{Ryzhik}
\be
\sum_{k}\BB{\frac{\sin^2(k/2)}{\sinh(z)}+\frac{\tanh(z/2)}{2}}^{-1} =
N\coth\B{\frac{Nz}{2}}-\coth\B{\frac{z}{2}}, \ k =  \frac{2\pi}{N},
\frac{4\pi}{N},\dots,\pi -\frac{\pi}{N}.
\label{ryzhik}
\ee
We multiply both sides of Eq. (\ref{ryzhik})  by $\tanh(z/2)$ and take the
derivative of the resulting equation with respect to $z$
\be
\sum_{k}\frac{\sin^2(k/2)}{\BB{\sinh^2(z/2)+\sin^2(k/2)}^2} = 
\frac{N}{\sinh(z)}\frac{d}{dz}\BB{\coth\B{\frac{Nz}{2}}\tanh\B{\frac{z}{2}}}.
\label{next}
\ee
Then, we multiply Eq. (\ref{next}) by $\cosh^4(z/2)$ and 
differentiate the resulting equation with respect to $z$ to get 
\be
\frac{d}{dz}\sum_{k} \frac{\sin^2k}{\BB{\sinh^2(z/2)+\sin^2(k/2)}^2}  =
\frac{2N}{\cosh^2\B{z/2}}\frac{d}{dz}\B{\frac{\cosh^3\B{z/2}}{\sinh\B{z/2}}
\frac{d}{dz}\BB{\coth\B{\frac{Nz}{2}}\tanh\B{\frac{z}{2}}}
}.
\label{next2}
\ee
Finally, we integrate Eq. (\ref{next2})  over $z$ from $0$ to $x$, 
use Eq. (\ref{next}) in the limit of $z\to0$ to compute 
$\sum_{k}\sin^{-2}(k/2)=(N^2-1)/6$, and rearrange the terms to obtain
\be
\sum_{k} 
\frac{\sin^2k}{\BB{\sinh^2(x/2)+\sin^2(k/2)}^2} =2N\B{\frac{\coth(Nx/2)}{\tanh(x)}-1} -\frac{N^2}{\sinh^2(Nx/2)}.
\label{moj}
\ee
Eq. (\ref{chinegative}) follows from the substitution of $x=\ln g$ into Eq. (\ref{moj}).

\begin{figure}[t]
\includegraphics[width=9.369cm]{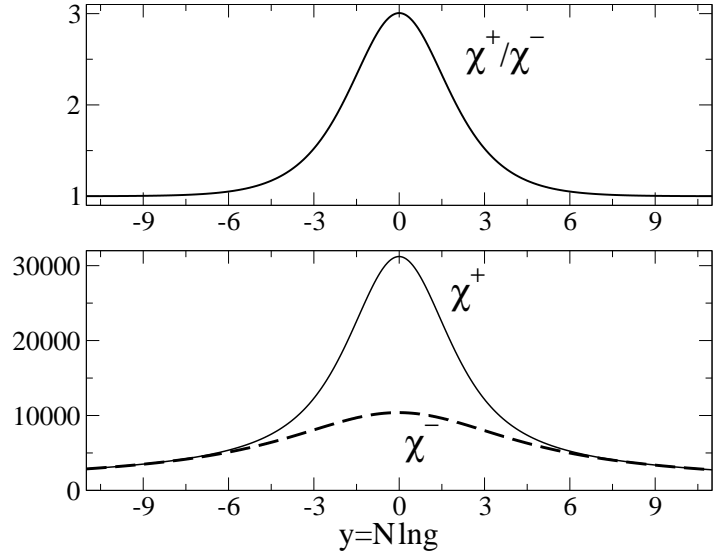}
\caption{Illustration of the difference between fidelity susceptibility of the positive and negative parity 
ground states. The former, i.e., $\chi^+(g)$, is given by Eq. (\ref{chipositive}). The
latter, i.e., $\chi^-(g)$ is given  by Eq. (\ref{chinegative}). 
Upper panel: $\chi^+/\chi^-$ as a function of 
$y=N\ln g\sim {\rm sign}(g-1)N/\xi(g)$, where $\xi(g)$ is the 
correlation length of the infinite Ising chain. Lower panel: 
$\chi^+$ (solid  line) vs. $\chi^-$ (dashed line). Both panels are  for $N=1000$. 
Near the critical point, $|y|<1$, $\chi^+$  differs from $\chi^-$ significantly. Far away from the critical point, 
$|y|\gg1$, the two curves perfectly merge. 
The system is in the ferromagnetic
(paramagnetic) phase for $y<0$ ($y>0$).}
\label{fig1}
\end{figure}

\section{Discussion}
\label{sec_discussion}
We will  discuss here the importance of proper quantization of
momenta in fidelity susceptibility calculations.  
If one evaluates fidelity susceptibility (\ref{chi}) over momenta (\ref{kpos}) 
then it equals $\chi^+(g)$. The evaluation of  sum 
(\ref{chi}) over momenta (\ref{kneg}) leads to  $\chi^-(g)$. The two
expressions are different.
Their   difference comes from shifting the summation over momenta by mere $\pi/N$,
which may seem to be insignificant  for $N\gg1$. This
impression is incorrect  near the critical point and correct  far away from it.
Two problems  arise from the improper quantization of momenta. 

{\bf Wrong symmetry of fidelity susceptibility in odd-sized systems}.
Fidelity susceptibility of the Ising chain  in the ground
state should be symmetric with respect to the 
\be
g \leftrightarrow -g
\label{gmg}
\ee
mapping  because the spin-spin interactions in the Hamiltonian (\ref{HIsing}) order the 
system in the $\pm x$ direction and so it should not matter  whether the magnetic
field points in the $+z$ or $-z$ direction. 

For even-sized systems the quantization of momenta (parity of the ground state) does not have to 
change after reflection (\ref{gmg}) to ensure the symmetry of fidelity susceptibility: 
$\chi^+(g)=\chi^+(-g)$ when $N$ is even. 

For odd-sized systems the situation is more complicated because neither $\chi^+(g)$ nor $\chi^-(g)$ is 
symmetric with respect to the mapping  (\ref{gmg}). 
Thus, one needs a different quantization of momenta (parity of the
ground state) for positive
and negative magnetic fields to ensure the symmetry of
fidelity susceptibility (Table \ref{table}). 
When properly calculated, ground state fidelity susceptibility is symmetric with respect to 
reflection (\ref{gmg}) in odd-sized systems due to the 
\bee
\chi^+(g)=\chi^-(-g)
\eee
relation. 
The wrong symmetry of fidelity susceptibility implies its wrong
magnitude near {\it at least} one of the critical points, which brings us to the second problem.

{\bf Wrong magnitude of fidelity susceptibility near the critical point}.
We need to define first what we mean by near and away from the
critical point. 
We express the distance from the critical point through the parameter
$y=N\ln g\sim {\rm sgn}(g-1)N/\xi(g)$ (we assume $g>0$ from now on). 
We say that the system is near the critical point when $|y|<1$ and away from it when $|y|>1$. 

To quantify for $g>0$ the difference between the correct (incorrect) 
expression for  ground state fidelity susceptibility 
given by $\chi^+$ ($\chi^-$), 
we study the ratio between the two expressions. As discussed
in  Appendix \ref{app_plusminus}, this ratio is unchanged by duality mapping  (\ref{KW}) and
it has a maximum at the critical point, where 
\bee
\left.\frac{\chi^+}{\chi^-}\right|_{y=0} = 3 \frac{N}{N-2}.
\eee
In other words, $\chi^+$ is roughly larger than $\chi^-$ by the factor of three near the critical
point (Fig. \ref{fig1}; see also Fig. 2 of Ref. \cite{ABQ2010}, where a similar plot is presented). 
This discrepancy happens even when $N\gg1$.
The ratio of $\chi^+/\chi^-$ monotonically goes to unity away from the
critical point
(Appendix \ref{app_plusminus} and Fig. \ref{fig1}), and one can easily verify that for $|y|\gg\ln N$
\bee
\frac{\chi^+}{\chi^-} -1 \simeq
4\exp\B{-|y|+\frac{|y|}{N}}\BB{N\sinh\B{\frac{|y|}{N}}-\cosh\B{\frac{|y|}{N}}}.
\eee
Therefore, far away from the critical point the difference between the correct  and incorrect
expression for fidelity susceptibility becomes negligible.

Both problems can be found in the literature. 
We will focus below on  popular Refs. \cite{Zanardi,GeometricXY,Chen2008}. 
In Ref. \cite{Zanardi} the system size is odd and momenta (\ref{kneg}) are used
for both positive and negative magnetic fields. 
The  expression for fidelity susceptibility derived in Ref. \cite{Zanardi} 
does not preserve the  $g\leftrightarrow-g$  symmetry \cite{remark_pre}.
The same problem appears in Ref. \cite{GeometricXY} (see Ref. \cite{remark_zanardi} for a brief discussion). 
In Ref. \cite{Chen2008}, the system size is  odd,  positive magnetic
fields are considered, and momenta (\ref{kneg}) instead of (\ref{kpos}) are 
used to compute fidelity susceptibility. Thus, the magnitude of fidelity susceptibility 
in Fig. 2 of Ref. \cite{Chen2008} is underestimated  around the  critical point 
by the factor of about three \cite{remark_wang}.

Summarizing, we have (i) comprehensively discussed  the ground state of the quantum Ising
model in a transverse field; (ii) computed and analyzed an exact  closed-form expression 
for the energy gap between the positive and negative parity eigenstates of the
Ising chain; (iii) calculated an exact  closed-form expression for fidelity susceptibility of the
Ising chain for all system sizes and magnetic fields; and (iv) discussed the
consequences of the incorrect quantization of momenta in fidelity susceptibility 
studies. 

The (i -- ii) results analytically illustrate the symmetry breaking
phenomenon in the Ising model, i.e., the disappearance of the energy gap leading to the degeneracy 
of the ground state. We show rigorously that the
gap decays exponentially fast with the system size in the thermodynamic limit.
The (iii) result completes the former studies of 
fidelity susceptibility  providing a textbook-quality expression. It
analytically shows how the fidelity susceptibility approach to quantum phase transitions 
works in the Ising model. 
The (iv) discussion corrects  misunderstanding in some former calculations
resulting from incorrect identification of the parity of the ground state. 
This misunderstanding leads to a  wrong expression for fidelity susceptibility near the critical point.

Finally, we mention two additional applications of our results. First, we expect that our exact analytical summation technique
will be useful in the future studies of various concepts in the  Ising model: The harmonic
oscillator of many-body physics. Second, our results for
fidelity susceptibility can be used to benchmark numerical results, e.g., the Quantum Monte Carlo simulations, where 
the complications arise due to the very existence of the two parity subspaces
that we discuss \cite{ABQ2010}. 
\\

The work of BD is supported by the Polish National Science Center grant DEC-2012/04/A/ST2/00088.
MMR acknowledges support from the FWF SFB grant F4014, 
 valuable discussions with Jacek Dziarmaga, and support
 through DEC-2011/01/B/ST3/00512.
 BD thanks Adolfo del Campo for  enjoyable  comments about the manuscript. 
\appendix

\section{Gap}
\label{appendix_gap}
We will prove here some properties of the gap mentioned in Sec. \ref{sec_ground}.
We will focus on the ferromagnetic phase, because the results for the
paramagnetic phase gap can be obtained through duality mapping (\ref{dual}).
Moreover, we will assume that the magnetic field $g>0$,
because the gap  has a well-defined parity with 
respect to the $g\leftrightarrow-g$ reflection. This leaves us with 
$g\in(0,1)$, which we assume through this section. 

We will use below two textbook inequalities. The first one comes from
Ref. \cite{Hardy1}
\be
a^r - b^r > r b^{r-1}(a-b) \for r>1   \  \vee \ r<0,
\label{ine}
\ee
where $a,b>0$ (equality happens when $a=b$, $r=0$, or $r=1$). The second one
is the Wendel's  inequality \cite{Wendel}
\be
\B{\frac{x}{x+s}}^{1-s} < \frac{\Gamma(x+s)}{x^s\Gamma(x)} < 1 \for s\in(0,1) \ \wedge \ x\in(0,\infty).
\label{Wendel}
\ee

{\bf Monotonicity of the gap}.
We will now prove that the gap increases monotonically from $g=0$ to $g=1$.
Since 
\bee
\frac{d}{dg}\B{\varepsilon^- - \varepsilon^+}=\int_0^1 dt\,\frac{4N}{\pi}
t^{N-3/2}\sqrt{\frac{1-t}{1-g^2t}} \frac{g^{N-1}}{\BB{1-(gt)^{2N}}^2}f(g,t),
\eee
it  suffices to show that $f(g,t)>0$:
\bee
\begin{aligned}
f(g,t) = &\BB{1-(gt)^{2N}}\BB{N-(N+1)g^2t} + 2N(1-g^2t)(gt)^{2N}\\
         &\ge 2N(gt)^{2N}\BB{1-g+N\B{\frac{1}{gt}-1-g+g^2t}}\\
	 &\ge 2N(gt)^{2N}(1-g)\BB{1+N(1-gt)} >0 \ \ {\rm for} \ \ t\in(0,1) \ \wedge \  g\in(0,1).
\end{aligned}
\eee
The first inequality follows from bounding of $1-(gt)^{2N}$ with inequality (\ref{ine}), where 
we substituted $a=1$, $b=gt$, and $r=2N$. The second one  results from
bounding $1/gt - 1$ with inequality (\ref{ine}), where now  $a=gt$,
$b=1$, and $r=-1$.

{\bf Lower bound on the gap}. 
\bee
\begin{aligned}
\varepsilon^--\varepsilon^+ = 
     &-2N\B{a_N+a_{3N}+a_{5N}+\dots} > -2Na_N\\
     &\ge 2N{\rm max}\B{\frac{\sqrt{1-g^2}}{\sqrt{\pi}}\frac{\Gamma(N-1/2)}{\Gamma(N+1)}g^N
     , \frac{2}{\pi} \frac{g^{N+1}}{N^2-1/4}}\\
     &> \max \B{ g^N \frac{2}{\sqrt{\pi}}\frac{\sqrt{1-g^2}}{\sqrt{N}} , g^{N} \frac{4g}{\pi N}}
\end{aligned}
\eee
The first inequality holds because $a_\ell<0$ (\ref{al}). The second follows 
from bounding  integral (\ref{al}) with  inequality 
\bee
 \sqrt{(1-t)(1-g^2 t)} = \sqrt{(1-g^2)(1-t)+g^2(1-t)^2} \ge \max\B{\sqrt{1-g^2}\sqrt{1-t},g(1-t)}.
\eee
The final step employs $1/(N^2-1/2)>1/N^2$ and $\Gamma(N-1/2)/\Gamma(N+1)>1/N^{3/2}$ following 
from inequality (\ref{Wendel}).

{\bf Upper bound on the gap}.  
\bee
\begin{aligned}
\varepsilon^--\varepsilon^+ = 
     &-2N\sum_{\ell=N,3N,\dots}a_\ell \\
     &\le \frac{4N}{\pi} g^{N+1} \sum_{\ell=N,3N,\dots}\frac{g^{\ell-N}}{\ell^2-1/4}   + \frac{2N}{\sqrt{\pi}}g^N\sqrt{1-g^2}\sum_{\ell=N,3N,\dots}\frac{\Gamma(\ell-1/2)}{\Gamma(\ell+1)}g^{\ell-N} \\
     &< \frac{4N}{\pi} g^{N+1}\frac{N}{N-1/2} \sum_{\ell=N,3N,\dots}\frac{g^{\ell-N}}{\ell^2}   + 
        \frac{2N}{\sqrt{\pi}}g^N\sqrt{1-g^2}\sqrt{\frac{N}{N-1}}
	  \sum_{\ell=N,3N,\dots} \frac{g^{\ell-N}}{\ell^{3/2}} \\
     &<g^{N} \frac{\pi g}{2N-1} + g^N 2 \frac{\sqrt{1 - g^2}}{\sqrt{N-1}} .
\end{aligned}
\eee
The first inequality follows from the substitution of 
\bee
\sqrt{(1-t)(1-g^2 t)} = \sqrt{(1-g^2)(1-t)+g^2(1-t)^2} \le \sqrt{1-g^2}\sqrt{1-t}+g(1-t)
\eee
into integral (\ref{al}). The second inequality follows from 
\bee
\frac{1}{\ell^2-1/4} < \frac{1}{\ell^2} \frac{2N}{2N-1}, \ \ 
\frac{\Gamma(\ell-1/2)}{\Gamma(\ell+1)}<\frac{1}{\ell^{3/2}}\sqrt{\frac{N}{N-1}},
\eee
both valid for $\ell\ge N$ [the latter follows from inequality (\ref{Wendel})].
In the final step, the sums are evaluated  at $g=1$.  

\section{$\chi^+/\chi^-$}
\label{app_plusminus}
We will discuss here the properties of $\chi^+/\chi^-$, which were only
briefly mentioned in Sec. \ref{sec_discussion}. Without loss of generality, 
we  consider magnetic fields $g>0$.

{\bf Duality of the ratio}. It can be shown using Eqs. (\ref{chipositive}) and (\ref{chinegative}) 
that 
\be
\frac{\chi^+(g)}{\chi^-(g)} =\frac{\chi^+(1/g)}{\chi^-(1/g)}.
\label{qwerty}
\ee
This ferromagnetic--paramagnetic  duality implies that 
we can focus  on the studies of $\chi^+/\chi^-$ in one of the phases. 
Symmetry (\ref{qwerty}) 
should not be taken for granted, because  neither $\chi^+$ nor $\chi^-$ is symmetric with
respect to the $g\leftrightarrow1/g$  mapping  (see Ref. \cite{BDfid3}
for the detailed discussion of the symmetries and properties of $\chi^+$).
Thus, we will further
restrict the discussion to $g\ge1$. The duality symmetry  also implies
that there is an extremum of $\chi^+/\chi^-$ at the critical point $g_c=1$.
Indeed, the  calculation of the first and second derivative of
$\chi^+/\chi^-$ shows that there is a maximum at $g_c=1$ for $N>2$. Note that
neither $\chi^+$ nor $\chi^-$ has maximum exactly at the critical point.

{\bf Monotonicity of the ratio}.
From now on, it is convenient to parameterize the ratio with $y=N\ln g$:
\be
\frac{\chi^+}{\chi^-} =
\tanh^2\B{\frac{y}{2}}\frac{\sinh(y-y/N)+(N-1)\sinh(y/N)}{\sinh(y-y/N)-(N-1)\sinh(y/N)}.
\label{qazwsx}
\ee
To prove that the ratio monotonically decreases in $y\in(0,\infty)$, 
we investigate its derivative
\bee
\begin{aligned}
&\frac{d}{dy}\B{\frac{\chi^+}{\chi^-}}=\frac{\tanh(y/2)}{\cosh^2(y/2)}
\frac{f(y)}{2\BB{\sinh(y-y/N)-(N-1)\sinh(y/N)}^2},\\
&f(y) = (N+1)\sinh^2\B{y-\frac{y}{N}} - (N-1)(N-2)\sinh^2(y)/N-(2N-1)(N-1)\sinh^2\B{\frac{y}{N}}.
\end{aligned}
\eee
Next, we  Taylor-expand the hyperbolic functions to get 
\bee
f(y) = -\frac{1}{2}\sum_{k=3}^\infty 
\frac{(2y)^{2k}}{(2k)!}\frac{N-1}{N^{2k}}a_k,\ \ a_k=(N-2)N^{2k-1} -(N+1)(N-1)^{2k-1}+2N-1.
\eee
Finally, one can prove by induction that $a_k>0$ for all $N>2$ and $k\ge3$.
Indeed, $a_3 = 2N(N^2-1)(N-2)(N-1/2)>0$  and if $a_k>0$ then 
$a_{k+1}>(2N-1)\B{N^2-1}\BB{(N-1)^{2k-2}-1}>0$. Therefore, $f(y>0)<0$ 
and so $d(\chi^+/\chi^-)/dy<0$ for $y>0$ and $N>2$, which we wanted to show. 

Calculating the ratio at $g=0,1,\infty$ and taking into account the
$y\leftrightarrow-y$ symmetry of Eq. (\ref{qazwsx}), we see that $\chi^+/\chi^-$ 
is monotonically decreasing from $3N/(N-2)$ at $g=1$ to unity  at $g=0$ and
$g=\infty$.


\begin{thebibliography}{34}
\bibitem[{rem({\natexlab{a}})}]{remark_pre}
\bibinfo{note}{For example, one can verify that the discrete sum for
  $S^\lambda_N(\lambda,\gamma=1)$ -- Eqs. (7) and (8) of Ref. \cite{Zanardi} --
  is equal to $4\chi^-(\lambda)$, whose asymmetry for {\it odd} $N$ with
  respect to the $\lambda\leftrightarrow-\lambda$ transformation is easily seen
  from our Eq. (\ref{chinegative}).}
\bibitem[{rem({\natexlab{b}})}]{remark_zanardi}
\bibinfo{note}{After setting $\gamma=1$ in Eq. (6) of Ref. \cite{GeometricXY},
  the $g_{11}$ component of the metric tensor represents fidelity
  susceptibility of the Ising chain. The expression given for it right under
  Eq. (6) of Ref. \cite{GeometricXY} is not symmetric under the transformation
  $h\leftrightarrow-h$, which can be easily checked (e.g., by plotting it). In
  fact, the discrete sum for $g_{11}(h,\gamma=1)$ from Ref. \cite{GeometricXY}
  equals $\chi^-(h)$, whose lack of reflection symmetry for odd-sized systems
  is evident from our Eq. (\ref{chinegative}).}
\bibitem[{rem({\natexlab{c}})}]{remark_wang}
\bibinfo{note}{One can show using our results that the expression for fidelity
  susceptibility that is plotted in Fig. 2 of Ref. \cite{Chen2008} equals
  $\chi^-(\lambda)$, while the correct result is $\chi^+(\lambda)$.}
\end{thebibliography}

\end{document}